\documentstyle[aps,prl,epsf]{revtex}

\begin{document}
\newcommand\beq{\begin{equation}}
\newcommand\eeq{\end{equation}}
\newcommand\bea{\begin{eqnarray}}
\newcommand\eea{\end{eqnarray}}

\def\eps{\epsilon}
\newcommand{\ket}[1]{| #1 \rangle}
\newcommand{\bra}[1]{\langle #1 |}
\newcommand{\braket}[2]{\langle #1 | #2 \rangle}
\newcommand{\proj}[1]{| #1\rangle\!\langle #1 |}
\newcommand{\ba}{\begin{array}}
\newcommand{\ea}{\end{array}}
\newtheorem{theo}{Theorem}
\newtheorem{defi}{Definition}
\newtheorem{lem}{Lemma}
\newtheorem{exam}{Example}
\newtheorem{prop}{Property}
\newtheorem{propo}{Proposition}

\twocolumn[\hsize\textwidth\columnwidth\hsize\csname
@twocolumnfalse\endcsname

\author{Barbara M. Terhal$^1$ and Pawe\l{} Horodecki$^2$,$^3$}

\title{A Schmidt number for density matrices}

 \address{\vspace*{1.2ex}
            \hspace*{0.5ex}{$^1$IBM Watson Research Center,
P.O. Box 218, Yorktown Heights, NY 10598, US;\,$^2$ 
Faculty of Applied Physics and Mathematics
Technical University of Gda\'nsk, Poland;\,
$^3$ Institute of Theoretical Physics, 
University of Hannover, D-30167 Hannover, Germany
}\\
Email: {\tt terhal@watson.ibm.com, pawel@mif.pg.gda.pl}}

\date{\today}

\maketitle
\begin{abstract}
We introduce the notion of a Schmidt number of a bipartite density matrix, 
characterizing the minimum Schmidt rank of the pure states that are needed 
to construct the density matrix. We prove that Schmidt number is nonincreasing 
under local quantum operations and classical communication. We show that 
$k$-positive maps witness Schmidt number, in the same way that positive maps
witness entanglement. We show that the family of states which is made from mixing the completely mixed state 
and a maximally entangled
state have increasing Schmidt number depending on the amount of maximally 
entangled state that is mixed in. We show that Schmidt number 
{\it does not necessarily increase} 
when taking tensor copies of a density matrix $\rho$;
we give an example of a density matrix for which the Schmidt numbers of 
$\rho$ and $\rho \otimes \rho$ are both $2$. 
\end{abstract}


]

In quantum information theory the study of bipartite entanglement is of great importance. The usual scenario is one in which two parties, Alice and Bob, share a 
supply of $n$ pure or mixed states $\rho^{\otimes n}$ which they would 
like to convert by Local Operations and Classical Communication (denoted as $LO+CC$) to 
a supply of $k$ other mixed or pure states $\sigma^{\otimes k}$, where $k$ 
can either be smaller or larger than $n$. 
The simple question that underlies many studies in bipartite entanglement 
is the question: what properties of these two sets of states make it 
possible or impossible to carry out such a protocol? Much work has been 
devoted to developing the necessary and sufficient conditions for this 
$LO+CC$ convertability. In the case of pure state convertability, it has 
been found that some aspects of this problem can be understood with 
the mathematics of majorization \cite{nielsen:major}. In the case of 
mixed state entanglement the theory of positive maps has been shown to 
play an important role \cite{nec_horo}. 
The power of positive maps is best illustrated 
by the Peres separability condition \cite{Asher96} which says that a 
bipartite density matrix which is unentangled (aka separable) 
must be positive under the application of the partial transposition map. 
For low dimensional spin systems this condition is not only necessary 
but also sufficient to ensure separability \cite{nec_horo}. It has been 
shown \cite{pptnodist} that density matrices which are positive under partial 
transposition are undistillable, that is, nonconvertible by $LO+CC$ to sets of 
entangled pure states. Many examples of these bound entangled states have been found \cite{bepawel,upb1,upb2,be:brussperes}. Evidence has been found as well
for the nondistillability of certain classes of entangled states which are 
not positive under partial transposition \cite{nptnond1,nptnond2}, and it was shown that this feature relates to the $2$-positivity of certain maps \cite{nptnond1}.

In this paper, we extend the $LO+CC$ classification of bipartite mixed states 
with the use of positive maps. In particular, we extend the notion 
of the Schmidt rank of a pure bipartite state to the 
domain of bipartite density matrices. We will show that this new quantity, 
which we will call {\em Schmidt number}, is witnessed by $k$-positive maps.

For a bipartite pure state which we write in its Schmidt decomposition 
(see Ref. \cite{peresbook}) 
\beq
\ket{\psi}=\sum_{i=1}^{k} \sqrt{\lambda_i} \ket{a_i} \otimes \ket{b_i}, 
\label{schmidt}
\eeq 
the number $k$ is the Schmidt rank of the pure state; it is the rank of 
the reduced density matrix $\rho_{red}={\rm Tr}_B \ket{\psi}\bra{\psi}$.
A necessary condition for a pure state to be convertible by $LO+CC$ 
to another pure state, is that the Schmidt rank of the first pure state 
is larger than or equal to the Schmidt rank of the latter pure state; 
local operations and classical communication cannot increase the Schmidt rank of a state \cite{lopop}.
When we extend this number to the domain of mixed states, we will require 
that this property of no increase of Schmidt number by $LO+CC$ still holds. 
We propose the following definition which is a natural
extension of the one applied to pure states.

\begin{defi}
A bipartite density matrix $\rho$ has Schmidt number $k$ if 
(i) for any decomposition of $\rho$, $\{p_i \geq 0, \ket{\psi_i}\}$ with 
$\rho=\sum_i p_i \ket{\psi_i}\bra{\psi_i}$ at least one of vectors 
$\{\ket{\psi_i}\}$ has at least Schmidt rank $k$ and (ii) there exists a decomposition 
of $\rho$ with all vectors $\{\ket{\psi_i}\}$ of Schmidt rank at most $k$.
\end{defi}

The Schmidt number of a pure state $\ket{\psi}$ is simply the Schmidt rank of 
the pure state. The Schmidt number of a separable state is 1. Let us denote
the set of density matrices on ${\cal H}_n \otimes {\cal H}_n$ that have 
Schmidt number $k$ or less by $S_k$. The set $S_k$ is a convex compact subset of the entire set of density matrices denoted by $S$, and $S_{k-1} \subset S_k$. The set of separable density matrices is $S_1$. 
 
The set $S_1$ has been completely characterized by positive
maps \cite{nec_horo}. Namely, for any state $\rho$ defined on ${\cal H}_n \otimes {\cal H}_n$, $\rho \in S_1$ holds if and only if the matrix $({\bf 1} \otimes
\Lambda_1) (\rho)$ has nonnegative eigenvalues for all positive maps 
$\Lambda_1: {\cal M}_{n}({\cal C}) \rightarrow{\cal M}_{n}({\cal C})$
\cite{maps}. 

Now let us recall the definition of $k$-positive linear maps:

\begin{defi}
The linear Hermiticity-preserving map $\Lambda$ is $k$-positive if and 
only if 
\beq
({\bf 1} \otimes \Lambda)(\ket{\psi}\bra{\psi}) \geq 0,
\label{posk}
\eeq
for all $\ket{\psi}\bra{\psi} \in S_k$. 
\end{defi}

It is not hard to show that when a positive map $\Lambda_n: 
{\cal M}_{n}({\cal C}) \rightarrow{\cal M}_{m}({\cal C})$ is $n$-positive,
$\Lambda_n$ is completely 
positive \cite{choi:can}. Similarly as with the characterization of $S_1$ in terms of 
$1$-positive (or, equivalently, positive) maps, we can characterize $S_k$ with $k$-positive maps:

\begin{theo}
Let $\rho$ be a density matrix on ${\cal H}_n \otimes {\cal H}_n$. The density matrix 
$\rho$ has Schmidt number at least $k+1$ if and only if there exists a $k$-positive linear map $\Lambda_k$ such that 
\beq
({\bf 1} \otimes \Lambda_k)(\rho) \not \geq 0.
\label{kwiteq}
\eeq
\label{kwitness}
\end{theo}

The proof of this theorem, which involves some technical details, is given at the 
end of this paper. With our definition of Schmidt number, it is not hard 
to prove that  

\begin{propo}
The Schmidt number of a density matrix cannot increase under local quantum 
operations and classical communication.
\label{inconv}
\end{propo}

{\em Proof}
Consider a density matrix $\varrho$ which has 
some Schmidt number $k$. Then it has the form  $\rho=\sum_i p_i
\ket{\psi_i}\bra{\psi_i}$ with all vectors $\ket{\psi_i}$ having Schmidt rank at 
most $k$. If there were any $LO+CC$ operation which would increase 
the Schmidt number of the state, it would increase
the Schmidt rank of least one of the pure states $\ket{\psi_i}\bra{\psi_i}$.
But no $LO+CC$ operation can increase the Schmidt rank of a pure state \cite{lopop}.
$\Box$

We will study a well known class of states $\rho_F$, mixtures of the 
completely mixed state and a maximally entangled state, by which we illustrate the 
notion of Schmidt number and its relation to $k$-positive maps.
First we note the following:

\begin{lem}
For any density matrix $\rho$ on ${\cal H}_N \otimes {\cal H}_N$ that has Schmidt number $k$, we have
\beq
f(\rho)\equiv \max_{\Psi} \bra{\Psi}\,\rho \,\ket{\Psi} \leq \frac{k}{N},
\eeq
\label{fef}
where we maximize over maximally entangled states $\ket{\Psi}$.
\label{boundf}
\end{lem}

{\em Proof} For any pure state $|\psi \rangle \langle \psi|$ with Schmidt rank 
$k$ characterized by its Schmidt coefficients $\{ \lambda_i \}$, 
see Eq. (\ref{schmidt}), the function $f$ equals \cite{filterhor}
\beq
f(|\psi \rangle \langle \psi|)= \frac{1}{N}\left[\sum_{i=1}^N \sqrt{\lambda_i}\right]^2.
\eeq
Using Lagrange multipliers to implement the constraint $\sum_i \lambda_i=1$ 
one can show that $[\sum_{i=1}^N \sqrt{\lambda_i}]^2 \leq k$. Since $\rho$ has 
Schmidt number $k$, $f(\rho)=\max_{\Psi} \sum_i p_i \langle \Psi |\psi_i \rangle \langle \psi_i |\Psi \rangle \leq
\frac{k}{N}$. $\Box$

We consider the family of states 
\bea
\rho_{F}=\frac{1-F}{N^2-1}\left({\bf 1}-\ket{\Psi^+}\bra{\Psi^+}\right)+
F \ket{\Psi^+}\bra{\Psi^+}, \nonumber \\
\ 0\leq F \leq 1,
\eea
with $\ket{\Psi^+}=\frac{1}{\sqrt{N}} \sum_{i=1}^N \ket{ii}$. When 
$F \leq \frac{1}{N}$ the density matrix $\rho_{F}$ is separable while 
for $F > \frac{1}{N}$ it can be distilled by an explicit protocol (see Ref. \cite{filterhor}). 

For these states we have $f(\rho_F)=F$. Therefore by Lemma \ref{boundf}, when $F > \frac{k}{N}$ the state has Schmidt number at least $k+1$.  
This result has an alternative derivation in terms of $k$-positive maps. There exists a well known family of $k$-positive maps for which the 
following has been proved:

\begin{lem} \cite{uhlpriv,tomiyama}
Let $\Lambda_p$ be a family of positive maps on ${\cal M}_{N}({\cal C})$
of the form
\beq
\Lambda_p(X)={\rm Tr}\, X {\bf 1}-p X,
\label{pmaps}
\eeq
where $X \in {\cal M}_{N}$. The map $\Lambda_p$ is $k$-positive , 
but $k+1$-negative ($k < N$) for 
\beq
\frac{1}{k+1} < p \leq \frac{1}{k}.
\eeq
\label{lempmap}
\end{lem}

Note that the range of $p$ for $k$-positivity does not depend on the 
dimension $N$ of ${\cal H}_N$. We note that the map $\Lambda_{p=1}$ 
is the reduction criterion that was used in Ref. \cite{filterhor} to develop a 
distillation method for entangled density matrices on ${\cal H}_N \otimes {\cal H}_N$.
If we apply these maps $\Lambda_p$ on half of $\rho_F$, we find the same lower 
bound on the Schmidt number of $\rho_F$. 

By giving an explicit decomposition of $\rho_F$, we will show that this 
lower bound on the Schmidt number is tight. We will do so by 
showing that the density matrix $\rho_{F}$ at the point $F=\frac{k}{N}$ 
can be made by mixing Schmidt rank $k$ vectors. If we show that at $F=\frac{k}{N}$ the density matrix can be made by mixing 
Schmidt rank $k$ vectors, then it follows that at any $F < \frac{k}{N}$ 
only vectors with Schmidt rank $k$ are needed, as we can make these states
by mixing the completely mixed state ${\bf 1}$ with the density matrix $\rho_{F=k/N}$.
As observed in Ref. \cite{filterhor}, the states $\rho_{F}$ have the important 
property that they are invariant under the operation $U \otimes U^*$ for 
any unitary transformation $U$. We can define the $LO+CC$ superoperator ${\cal S}^{U \otimes U^*}$ as
\beq
{\cal S}^{U \otimes U^*}(\rho)=\frac{1}{Vol(U)}\int \, dU\, U \otimes U^* \rho\, U^{\dagger} \otimes {U^*}^{\dagger},
\label{defS}
\eeq
which will bring any initial state $\rho$ into the form of $\rho_F$, 
i.e. a mixture of ${\bf 1}$ and $\ket{\Psi^+}\bra{\Psi^+}$.

As our initial state we take the maximally entangled Schmidt rank $k$ state 
\beq
\ket{\psi_k}= \frac{1}{\sqrt{k}} \sum_{i=1}^{k}\ket{ii}, 
\eeq
and let ${\cal S}^{U \otimes U^*}$ operate on this state. We easily find 
that the resulting density matrix equals 
$\frac{k}{N}\ket{\Psi^+}\bra{\Psi^+}+\frac{1-k/N}{N^2-1}({\bf 1}-\ket{\Psi^+}\bra{\Psi^+})$ which is the desired result. We can summarize these results in a 
theorem:

\begin{theo}The state $\rho_F$ in ${\cal H}_N \otimes {\cal H}_N$ has 
Schmidt number $k$ if and only if 
\beq
\frac{k-1}{N} < F \leq \frac{k}{N}.
\eeq
\end{theo}

For this special class of states we have found that Schmidt number is 
monotonically related to the amount of entanglement in the state. 
This is not always the case; a pure state $\ket{\psi}$ 
with Schmidt rank $k$ can have much less entanglement than, say, the 
1 bit of a maximally entangled Schmidt rank 2 state. 

When we find that a density matrix $\rho$ is of Schmidt number $k$, 
we may ask whether the tensor product $\rho \otimes \rho$ is 
of Schmidt number $k^2$. Or is it possible to make $\rho \otimes \rho$ 
from Schmidt rank $m < k^2$ vectors? In other words, when 
we assign the value ${\cal N}(\rho)=\log k$ to a density matrix $\rho$ which has 
Schmidt number $k$, we ask whether ${\cal N}(\rho)$ is {\em additive}, i.e. 
\beq
{\cal N}(\rho^{\otimes n})=n \log k. 
\label{addi}
\eeq

For pure states this additivity property holds: The tensor product 
of two pure entangled states, each with Schmidt rank $k$, is a pure state 
with Schmidt rank $k^2$. With a simple argument we can lower bound the function ${\cal N}(\rho)$. 
When $\rho$ itself has Schmidt number $k$, then the Schmidt number of any number of copies of $\rho$, $\rho^{\otimes n}$, must be at least $k$.
We can get $\rho$ from $\rho^{\otimes n}$ ($n \geq 2$) by local operations, 
namely tracing out all states but one. Therefore $\rho^{\otimes n}$ cannot 
have a smaller Schmidt number than $\rho$ itself, since, if it were, then we 
find a contradiction with Proposition \ref{inconv}. Thus we obtain the bound 
\beq
{\cal N}(\rho^{\otimes n}) \geq {\cal N}(\rho).
\eeq
Let us consider tensor copies of the state $\rho_F$, i.e. 
$\rho_F^{\otimes m}$ with $m \geq 2$ and lower bound the Schmidt number as 
before. This time, in the space ${\cal H}_{N^m} \otimes {\cal H}_{N^m}$ we have $f(\rho_F^{\otimes m}) \geq F^m$ and with Lemma \ref{boundf} this implies that when $F^m > \frac{k}{N^m}$ the Schmidt number of $\rho_F^{\otimes m}$ is at 
least $k+1$. We will give an example of two copies of $\rho_F$ in ${\cal H}_2
\otimes {\cal H}_2$ which will show that this bound can be tight.
The idea is again to use the $U \otimes U^*$ invariance of each copy 
of $\rho_F$. We show how to construct the density matrix $\rho_F^{\otimes 2}$ where 
$\rho_F$ is a 2-qubit state at $F=\frac{1}{\sqrt{2}}$ by mixing Schmidt 
rank two vectors. Let $\ket{\psi}$ be the state 
\beq
\ket{\psi}=\frac{1}{\sqrt{2}} \left[\ket{0,\psi_0} \otimes \ket{0,\psi_0} +\ket{1,0} \otimes \ket{1,0}\right],
\eeq
where 
\beq
\ket{\psi_0}=\sqrt{2} \sqrt{\sqrt{2}-1}\ket{0}+(1-\sqrt{2})\ket{1}.
\eeq
This is a maximally entangled Schmidt rank two state between 
${\cal H}_{A_1,A_2}$ and ${\cal H}_{B_1,B_2}$. Now Alice and Bob perform 
the following operations on this state:
\begin{itemize}
\item They perform the superoperator ${\cal S}^{U \otimes U^*}$, 
Eq. (\ref{defS}), on the Hilbert space ${\cal H}_{A_1} \otimes {\cal H}_{B_1}$ and 
they apply the same superoperator on ${\cal H}_{A_2} \otimes {\cal H}_{B_2}$.
\item Then Alice and Bob symmetrize the state between system 1 and system 2, 
i.e. with probability 1/2, they locally swap qubit $A_1 \leftrightarrow A_2$ 
and $B_1 \leftrightarrow B_2$ and with probability 1/2 they do nothing.
\end{itemize}

By these two $LO+CC$ operations any initial state is mapped onto a state 
of the form 
\beq
\ba{l}
a\, ({\bf 1}-\ket{\Psi^+}\bra{\Psi^+})^{\otimes 2}+b\, \ket{\Psi^+}\bra{\Psi^+} \otimes[{\bf 1}-\ket{\Psi^+}\bra{\Psi^+}]\\
+b\,[{\bf 1}-\ket{\Psi^+}\bra{\Psi^+}] \otimes \ket{\Psi^+}\bra{\Psi^+}+c\, (\ket{\Psi^+}\bra{\Psi^+})^{\otimes 2}  
\ea
\eeq

\begin{figure}
\epsfxsize=7cm
\epsfbox{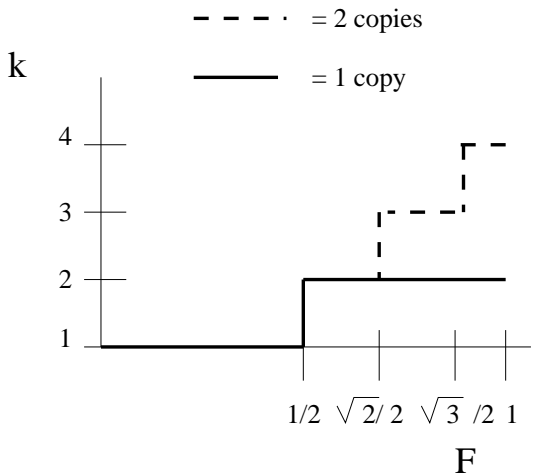}
\caption{The Schmidt number of one and two copies of $\rho_F$ for $N=2$ 
as a function of $F$.}
\label{stepk}
\end{figure}

By going through the algebra, one can check that the two operations on 
the state $\ket{\psi}$ result in $a=\frac{[\sqrt{2}-1]^2}{18}$, $b=\frac{\sqrt{2}-1}{6}$ and $c=\frac{1}{2}$. We have shown that the 
state $\rho_{F=\frac{1}{\sqrt{2}}}$ has Schmidt number two {\em and} 
$\rho_{F=\frac{1}{\sqrt{2}}}^{\otimes 2}$ has Schmidt number 2! This is an 
example of nonadditivity of the function ${\cal N}$ in Eq. (\ref{addi}).
We have numerical evidence that the lower bound on the Schmidt number at 
$F=\frac{\sqrt{3}}{2}$ is tight as well, i.e. it seems possible to make 
$\rho_F^{\otimes 2}$ at $F=\frac{\sqrt{3}}{2}$ with mixing only Schmidt rank 
3 vectors. The stepwise behavior of Schmidt number is illustrated in Figure 
\ref{stepk}.

Since we have found that the Schmidt number can exhibit `nonadditivity', 
we can define the asymptotic Schmidt characteristic of a state
\beq
{\cal N}^{\infty}(\rho)=\lim_{m \rightarrow \infty} \left\{\frac{\log k}{m}\; :\; 
\rho^{\otimes m} \mbox{ has Schmidt number } k\right\}.
\eeq 

The asymptotic Schmidt number of a density matrix gives us some information 
about whether two states, whose single copy Schmidt number is identical,
can be interconverted by $LO+CC$. Say we have two density 
matrices $\rho_1$ and $\rho_2$ which each have Schmidt number $k$. But assume that for some $n$ the Schmidt number of $\rho_1^{\otimes n}$ is larger than the Schmidt number of $\rho_2^{\otimes n}$. 
Then it follows that a single copy of $\rho_2$ cannot be converted by $LO+CC$
to $\rho_2$, because, if it could, then by repeating this procedure $n$ 
times we could convert $\rho_2^{\otimes n}$ to  $\rho_1^{\otimes n}$, which 
is in violation of Proposition \ref{inconv}.

In conclusion, we have introduced a new criterion for $LO+CC$ convertibility
for bipartite mixed states and shown its relation to $k$-positive maps. 
Since the theory of $k$-positive maps is not (yet) greatly developed, we 
have not been able to use this connection extensively. We have found that 
Schmidt number for mixed states behaves differently than for pure states, 
in particular it does not necessarily increase when taking tensor copies of a state. This feature of 'nonadditivity' makes it possible to pose the following 
open question: Assume that a bipartite state $\rho_1$ has a higher Schmidt 
number than $\rho_2$, and thus $\rho_2$ cannot be converted to $\rho_1$.
Assume however that for some $\rho_{help}$ we can prove that 
$\rho_1 \otimes \rho_{help}$ and $\rho_2 \otimes \rho_{help}$ have the 
{\em same} Schmidt number or that $\rho_2 \otimes \rho_{help}$ has a larger 
Schmidt number. Then it could possibly be that 
$\rho_2 \otimes \rho_{help}$ can be converted to $\rho_1 \otimes \rho_{help}$. 
This would be an example of the use of borrowing of entanglement for mixed state conversions. Such a borrowing scheme has been found for exact pure state conversion 
\cite{jonaplenio}; it would be interesting to see whether it is possible 
in the mixed state domain.

{\bf Acknowledgments}: BMT would like to thank David DiVincenzo and 
Armin Uhlmann for interesting discussions. BMT acknowledges support of 
the ARO under contract number DAAG-55-98-C-0041. PH is grateful to Maciej Lewenstein 
and Karol \.Zyczkowski for stimulating discussions. 
He also acknowledges support 
from Polish Committee for Scientific Research, 
grant No. 2 PO3B 103 16, and partial support from the Deutscher Akademischer
Austauschdienst. 
Part of this work has been carried out at the ESF-Newton Institute 
Workshop on `Computation, Complexity and the Physics of Information' held 
in Cambridge U.K., July 1999.

\vspace{0.5cm}
{\bf Proof of Theorem 1}

Before proving Theorem \ref{kwitness} it will be useful to give some properties 
related to $k$-positivity and an alternative formulation of $k$-positivity:

\begin{lem}
The linear Hermiticity-preserving map $\Lambda$ is $k$-positive if and only 
if 
\beq
({\bf 1} \otimes \Lambda)(\ket{\Psi_k}\bra{\Psi_k}) \geq 0,
\label{maxe_only}
\eeq
for all $\ket{\Psi_k}$ which are maximally entangled Schmidt rank $k$ vectors.
The linear Hermiticity-preserving map $\Lambda$ is $k$-positive if and only 
if 
\beq
\sum_{n,m=1}^k \sqrt{\mu_n \mu_m} \bra{b_n}\, \Lambda( \ket{a_n} \bra{a_m})\, \ket{b_m} \geq 0,
\label{reform}
\eeq
for all possible orthogonal sets of vectors $\{\ket{a_n}\}_{n=1}^k$ and $\{\ket{b_n}\}_{n=1}^k$ and Schmidt coefficients $\{\mu_n\}$, $\sum_{n=1}^k \mu_n=1$. 
Finally, if $\Lambda$ is $k$-positive, then $\Lambda^{\dagger}$ defined by ${\rm Tr}\, A^{\dagger} \Lambda(B)={\rm Tr}\, \Lambda^{\dagger}(A^{\dagger}) B$ for all $A$ and $B$, is also $k$-positive. 
\label{propkpos}
\end{lem}

{\em Proof (sketch)} 
Equation (\ref{maxe_only}) can be proved in a fashion completely analogous to Lemma 7 in Ref. \cite{nptnond1}; there it is proved for $k=2$. Equation (\ref{reform})
has been proved by A. Uhlmann \cite{uhlpriv} and can be understood as follows.
It is equivalent to 
\beq
\ba{c}
\sum_{n,m,i,j=1}^k \sqrt{\mu_n \mu_m}  \\
\bra{\gamma_n,b_n} \,
({\bf 1} \otimes \Lambda)(\ket{\gamma_i,a_i} \bra{\gamma_j,a_j})
\ket{\gamma_m,b_m} \geq 0, 
\ea
\eeq
for any orthogonal set $\{\ket{\gamma_n}\}_{n=1}^k$, or 
\beq
\bra{\psi}\, ({\bf 1} \otimes \Lambda)(\ket{\Psi_k} \bra{\Psi_k})\, \ket{\psi} \geq 0,
\eeq
for arbitrary $\ket{\psi}$ and $\ket{\Psi_k}$. The $k$-positivity of $\Lambda^{\dagger}$ can be seen by noting that Eq. (\ref{posk}) can be written in terms of $\Lambda^{\dagger}$ as 
\beq
\bra{\psi}\, ({\bf 1} \otimes \Lambda^{\dagger})(\ket{\phi} \bra{\phi})\, \ket{\psi} \geq 0,
\eeq
for all $\ket{\psi}\bra{\psi} \in S_k$ and arbritrary $\ket{\phi}$. Since $\ket{\psi}$ has Schmidt rank $k$ or less, it follows that we can restrict the state 
$\ket{\phi}$ to be of Schmidt rank $k$ as well, or $\ket{\phi}\bra{\phi} \in S_k$. $\Box$

{\em Proof of Theorem \ref{kwitness}}
Consider the ``if'' part of the theorem.
Suppose, conversely, that there exist some $\varrho$
of Schmidt number at most $k$ for which at some time
Eq. (\ref{kwiteq}) is satisfied for some k-positive map
$\Lambda_k$. The first assumption guarantees that $\varrho$
is a convex combination of pure states $| \psi_i \rangle \langle \psi_i|$
of Schmidt rank at most $k$ each. From the definition of $k$-positivity it
follows immediately that all $| \psi_i \rangle \langle \psi_i|$
remain positive after the action of ${\bf 1} \otimes \Lambda_k$, 
so their convex combination equal to $\varrho$ remains positive too.
But this then is in a contradiction with Eq. (\ref{kwiteq}). \\
Consider the ``only if'' part of the theorem. Let $\rho$ have Schmidt number at least $k+1$, i.e. $\rho \not \in S_k$.
Then there exists a hyperplane $\{\sigma \in S\,|\,{\rm Tr}\,H\sigma=0\}$
which separates the convex compact set $S_k$ and the point $\rho \not \in S_k$,
i.e. there exists a Hermitian operator $H$ such that
\beq
{\rm Tr}\, H \rho < 0\,\,\,{\rm and}\,\,\, \forall \sigma \in S_k\,\, {\rm Tr}\,H\sigma \geq 0.
\eeq 
We will show that we can associate with this Hermitian operator $H$ a 
positive linear map $\Lambda_k$ which is $k$-positive. We define $\Lambda$ as 
\beq
H=({\bf 1} \otimes \Lambda_k)(\ket{\Psi^+}\bra{\Psi^+}),
\label{hlam}
\eeq
where $\ket{\Psi^+}=\frac{1}{\sqrt{n}} \sum_{i=1}^n \ket{ii}$. We use ${\rm Tr}\,H\sigma \geq 0$ where we take $\sigma$ to be an entangled Schmidt rank $k$ vector, $\sigma=\sum_{n,m=1}^k \sqrt{\lambda_m \lambda_n} \ket{a_m,b_m} \bra{a_n,b_n}$. 
We will denote transposition in the full $\{\ket{a_n}\}$ basis as $T^a$. We can rewrite the expression 
\beq
\sum_{n,m=1}^k \sqrt{\lambda_m \lambda_n} {\rm Tr}\, H \,\ket{a_m,b_m} \bra{a_n,b_n} \geq 0,
\eeq
using Eq. (\ref{hlam}) and the expansion of a linear map ${\cal S}$ on an operator 
$X$ as ${\cal S}(X)=\sum_{i,j=1}^n \bra{i} X \ket{j} {\cal S}(\ket{i}\bra{j}) $, 
as 
\beq
\frac{1}{n} \sum_{m,n=1}^k \sqrt{\lambda_n \lambda_m} \bra{b_n}\,  \Lambda_k \circ T \circ T^a (\ket{a_n} \bra{a_m})\, \ket{b_m} \geq 0.  
\eeq
Lemma \ref{propkpos} then implies that the map $\Lambda_k \circ T \circ T^a$ is $k$-positive. The map $T \circ T^a$ maps the vector $\ket{a_n}$ onto $\ket{a_n^*}$ where 
complex conjugation is performed with respect to the $\{\ket{i}\}$ basis. This 
corresponds to a unitary rotation from the $\{\ket{a_n}\}$ basis to
the
$\{\ket{a_n^*}\}$ basis. Therefore $\Lambda_k$ itself will be $k$-positive. On the other 
hand the condition ${\rm Tr}\, H \rho < 0$ can be rewritten, using Eq. (\ref{hlam}), as 
\beq
\bra{\Psi^+}\,({\bf 1} \otimes \Lambda_k^{\dagger}) (\rho)\, \ket{\Psi^+} < 0.
\eeq
Since $\Lambda_k$ is $k$-positive, $\Lambda_k^{\dagger}$ is $k$-positive, see Lemma \ref{propkpos}. This completes the proof.
$\Box$


\bibliographystyle{hunsrt}
\bibliography{refs}

\begin{thebibliography}{10}
\bibitem{maps}
Let us recall that the linear map $\Lambda_1$ is said to be positive if it
  transforms any matrix with nonnegative eigenvalues into a matrix with
  nonnegative eigenvalues. ${\cal M}_{n}$ stands for the set of all complex $n
  \times n$ matrices.
\end{thebibliography}

\end{document}